\documentclass[twocolumn,superscriptaddress,secnumarabic,amssymb, nobibnotes, aps, bm,prb,floatfix,longbibliography]{revtex4-2}

\usepackage[english]{babel}
\usepackage[latin9]{inputenc}
\usepackage[usenames,dvipsnames]{xcolor}
\usepackage{amsmath}
\definecolor{goodgreen}{rgb}{0.1,0.5,0}
\definecolor{goodred}{rgb}{0.7,0,0}
\usepackage[colorlinks,urlcolor=goodgreen,citecolor=blue,linkcolor=goodred]{hyperref}

\newcommand{\cL}{{\cal L}}

\newcommand{\beq}{\begin{equation}}
\newcommand{\eeq}{\end{equation}}
\newcommand{\bea}{\begin{eqnarray}}
\newcommand\bal{\begin{aligned}}
\newcommand\eal{\end{aligned}}
\newcommand{\eea}{\end{eqnarray}}

\newcommand{\pr}{\partial}
\newcommand{\prdt}{\partial\cdot}

\newcommand{\mQ}{\mathcal{Q}}
\newcommand{\mS}{\mathcal{S}}
\newcommand{\mV}{\mathcal{V}}
\newcommand{\mU}{\mathcal{U}}

\begin{document}
\title{Massive Higher-Spin Fields in the Fractional Quantum Hall Effect}

\author{E. Bergshoeff}
\email{e.a.bergshoeff@rug.nl}
\affiliation{
Van Swinderen Institute, University of Groningen
Nijenborgh 4, 9747 AG Groningen, The Netherlands}

\author{A. Campoleoni}
\email{andrea.campoleoni@umons.ac.be}
\affiliation{Service de Physique de l'Univers, Champs et Gravitation, Universit\'e de Mons~--~UMONS, 20 place du Parc, 7000 Mons, Belgium}

\author{G. Palumbo}
\email{giandomenico.palumbo@gmail.com}
\affiliation{School of Theoretical Physics, Dublin Institute for Advanced Studies, 10 Burlington Road, Dublin 4, Ireland}

\author{P. Salgado-Rebolledo}
\email{psalgadoreb@hep.itp.tuwien.ac.at}
\affiliation{Institute for Theoretical Physics, TU Wien, Wiedner Hauptstr.~8-10/136, A-1040 Vienna, Austria}
\affiliation{Institute of Theoretical Physics,
Wroc\l{}aw University of Science and Technology, 50-370 Wroc\l{}aw, Poland}
\affiliation{Instituto de Ciencias Exactas y Naturales (ICEN), Universidad Arturo Prat, Playa Brava 3256, Iquique, Chile}

\begin{abstract}
\noindent Incompressibility plays a key role in the geometric description of fractional quantum Hall fluids. 
It is naturally related to quantum area-preserving diffeomorphisms and the underlying Girvin-MacDonald-Plazman algebra, which gives rise to an emergent non-relativistic massive spin-$2$ mode propagating in the bulk. 
The corresponding metric tensor can be identified with a nematic order parameter for the bulk states. 
In the linearized regime with a flat background, it has been shown that this mode can be described by a spin-$2$ Schroedinger action. 
However, quantum area-preserving diffeomorphisms also suggest the existence of higher-spin modes that cannot be described through nematic fractional quantum Hall states. 
Here, we consider p-atic Hall phases, in which the corresponding p-atic order parameters are related to higher-rank symmetric tensors. We then show that in this framework, non-relativistic massive chiral higher-spin fields naturally emerge and that their dynamics is described by higher-spin Schroedinger actions. 
We finally show that these effective actions can be derived from relativistic massive higher-spin theories in $2+1$ dimensions after taking a non-relativistic limit.
\end{abstract}
\date{\today}
\maketitle

\section{Introduction}

The notion of order parameter is useful as a unifying concept for describing states of matter with emergent macroscopic properties. 
For instance, superconductors \cite{Schrieffer} and liquid crystals \cite{deGennes} are characterized by local-order parameters, which reflect the presence of spontaneously broken symmetries. 
On the other hand, topological phases such as fractional quantum Hall (FQH) states are instead characterized by topological invariants. 
Consequently, they cannot be described within Landau's theory of phase transitions \cite{PachosBook, Wen, Wen2}.
However, there exist several states of matter where local order and topological phases can co-exist, such as topological superconductors \cite{Qi, Hasan, Kitaev, Lisle1, Lisle2, Maraner} and nematic FQH phases \cite{Dorsey, Mulligan, Maciejko, Fradkin3, You, Papic5, Regnault, Manfra, Pu}.
In the latter, nematicity can be induced, for instance, through a tilted magnetic field, which naturally stretches the Landau orbits while preserving the areas the orbits encircle.
The corresponding nematic order is associated to a rank-$2$ symmetric tensor order parameter that gives an emergent ``spin-$2$'' collective massive mode that appears in the bulk \cite{Haldane, Haldane2, Son5, Bradlyn, Gromov, Rezayi} and has been recently observed experimentally \cite{Pinczuk}. 
The importance of nematic order in this context can be qualitatively understood by noting that Hall nematic phases, which break rotational symmetry while preserving a $C_2$ discrete rotational symmetry, preserve the topological states while the corresponding deformations can be seen as special kinds of area-preserving deformations of the Hall fluids.
The corresponding emergent quantum (non-commutative) geometry is, in fact, invariant under quantum area-preserving diffeomorphisms \cite{Cappelli2, Karabali, Oblak3, Palumbo2023}, that encode the incompressibility of quantum Hall fluids, and consequently are responsible for the existence of an energy bulk gap.
These features are encoded in the so called Girvin-MacDonald-Plazman (GMP) algebra \cite{GMP}, also known as $W_\infty$ algebra  in the high-energy-physics literature \cite{Cappelli2}\footnote{The $W_\infty$ algebra mentioned here is a deformation of the algebra of area-preserving diffeomorphisms of a cylinder, called the $w_\infty$ algebra. This deformation is linear in the generators and was constructed in \cite{Pope:1989ew}. This infinite-dimensional algebra should be distinguished from (i) the 3D higher-spin algebra $hs(1,1)$ that was used in \cite{Bergshoeff:1989ns} to construct a gauge theory of 3D massless higher spins and is related to the algebra of area-preserving diffeomorphisms of the two-dimensional hyperbolic space and (ii) from the asymptotic symmetry algebra of the $hs(1,1)$ gauge theory which is a nonlinearly realized $W_\infty$ algebra \cite{Henneaux:2010xg} (see also \cite{Campoleoni:2010zq, Campoleoni:2014tfa}). For a one-parameter supersymmetric extension of the (linearly realized) $W_\infty$ algebra, see \cite{Bergshoeff:1991dz}.}. 
Thus, the emergent geometry in Hall fluids is associated to a non-relativistic massive graviton-like mode, which at linearised level around a flat background can be described by a spin-$2$ Schroedinger action \cite{Gromov}.
A characteristic feature of the GMP/$W_\infty$ algebra is that, besides a generator of conformal spin two, it also contains an infinite number of generators of increasing conformal spin, with each spin occurring once \cite{Bergshoeff:1989ns}. 
This suggests the existence of additional bosonic bulk modes of spin greater than two (aka higher-spin modes), and this expectation is supported by several theoretical and numerical evidences \cite{Son4, Cappelli, Liu, Lapa}. However, the quantum-field-theory characterization of these massive modes remains unclear.

In this work, we partially address this important open question by employing a complementary approach in which we show that these higher-spin modes can naturally appear in the fractional quantum Hall effect (FQHE) via p-atic phases, with p even such that inversion symmetry is preserved \cite{Haldane2023}. 
In p-atic liquid crystals, the corresponding p-atic order parameters are higher-rank traceless symmetric tensors, which in two spatial dimensions are related to $C_\textrm{p}$ symmetries,
namely p-fold rotational symmetries
\cite{Giomi2, Giomi3}. 
They are well known in soft matter and naturally generalize nematic phases \cite{deGennes, Nelson, Lubensky, Giomi, Zaanen, Zaanen2}.
Notice that although several p-atic Hall phases have been already proposed and studied \cite{Balents, Musaelian, Fradkin2, Fogler, Wexler}, their relation with higher-spin fields has been overlooked.
Here, we will show that within this framework, massive higher-spin fields naturally emerge and their dynamics is described by higher-spin Schroedinger actions near the isotropic phase transition. 
In particular, we will focus on the chiral and massive spin-$4$ mode, which has been numerically discussed in \cite{Liu} and that we will show to be related to a tetratic order parameter. We will also show, through symmetry and field-theory arguments, that these effective actions can be derived from relativistic massive higher-spin theories after taking a suitable non-relativistic limit. Finally, we will propose a measurement protocol to detect spin-4 modes in FQH states.

\section{P-atic phases in the FQHE}

In this section, firstly we summarize the effective-field-theory description of nematic order in the FQHE by mainly following \cite{Maciejko}. Secondly, we generalize those results to the case of tetratic Hall states. De Gennes' nematic order in liquid crystals is characterized by a rank-$2$ symmetric and traceless order parameter, which in two spatial dimensions takes the form $Q_{ab}({\bf{x}},t)=Q_N(2 d_a d_b-\delta_{ab})$, where ${\bf{d}}= ( \cos \vartheta/2, \sin \vartheta/2)$ is the planar nematic director \cite{Virga}.
Since it has only two independent components, we can define from it the complex order parameter
\begin{equation}\label{nematic}
	\mathcal{Q}_N=Q_{11}+i Q_{12}= Q_N e^{i \vartheta}.
\end{equation}
In the long-wavelength limit \cite{Wen2}, the standard topological Chern-Simons (CS) sector of the Laughlin states reads
\begin{equation}\label{Landau-FQHE-Laughlin}
\mathcal {L}_{CS}=  \frac{q}{4 \pi} \epsilon^{\mu\nu\lambda} a_{\mu}\partial_{\nu} a_{\lambda}-\frac{1}{2 \pi}
\epsilon^{\mu\nu\lambda} A_{\mu} \partial_{\nu} a_{\lambda},
\end{equation}
with $A_\mu$ and $a_\mu$ the electromagnetic and U(1) emergent gauge fields, respectively, and where $\mu, \nu \in \{0,1,2\}$. Here, $q$ is an odd integer number associated with the filling factor such that the Hall conductivity is quantized to $(1/q) (e^2/h)$. The corresponding effective action for the nematic Laughlin states is given by \cite{Maciejko}
\begin{eqnarray}\label{Landau-FQHE}
\mathcal{L}_{NFQHE}=\mathcal{L}_{CS}+\mathcal{L}_N
-\frac{1}{2 \pi}
\epsilon^{\mu\nu\lambda} a^N_{\mu} \partial_{\nu} a_{\lambda},\nonumber \\
a^N_c = - \frac{1}{8} \epsilon_{bd} Q_{ab} \partial_c Q_{ad}+ C \epsilon_{bd} \partial_b Q_{dc}, \nonumber \hspace{0.3cm}\\
a^N_0 = - \frac{1}{8} \epsilon_{bd} Q_{ab} \dot{Q}_{ad}, \hspace{1.4cm}
\end{eqnarray}
where $\dot Q_{ab}\equiv \partial Q_{ab}/\partial t$, $a^N_0$ can be associated with a Berry phase, while $C$ is a constant parameter (it is related to the equal-time structure factor as shown in \cite{Maciejko}). 
Close to the phase transition from an isotropic to a nematic FQH phase, $\mathcal{L}_N$ in the above equation identifies the effective Landau-de Gennes (LdG) Lagrangian
\begin{equation}\label{Landau}
\mathcal{L}_{N}= \frac{k_1}{8} \epsilon_{bc} Q_{ab}  \dot{Q}_{ac}- \frac{k_2}{8} (\partial_c Q_{ab})^2-\frac{k_3}{2} (Q_{ab})^2+\cdots ,
\end{equation}
where $k_i$ are some given physical parameters and we omitted higher powers of the nematic order parameter. The first term proportional to the Berry phase $a^N_0$ originates
from the broken time-reversal invariance of the FQHE and dictates the
quantum dynamics of the chiral non-relativistic ``graviton'' \cite{Rezayi}.
Importantly, we omitted higher-powers in $Q_{ab}$ because we consider the isotropic phase in presence of small nematic deformations.
The corresponding Hamiltonian
\begin{equation}\label{Hamiltonian}
H_{N}= \frac{k_2}{8} |\partial_i \mathcal{Q}_N|^2+\frac{k_3}{2} |\mathcal{Q}_N|^2 ,
\end{equation}
describes a quadratic energy spectrum that identifies a gapped spin-$2$ mode, namely the GMP mode.

We are now ready to discuss the generalization of the above results to the case of tetratic order. The latter refers to a generalization of nematic order, in which a fourfold rotational symmetry $C_4$ replaces the $C_2$ of nematic phases. 
This phase can be either considered as the result of a deformation of the magnetic field or of the confining potential preserving the $C_4$ symmetry, or as a result of local orientational ordering introduced by interactions, possibly supplemented by additional p-atic modes \cite{Balents, Musaelian, Fradkin2, Fogler, Wexler}. 
Being inversion invariant, tetratic phases are compatible with FQH states \cite{Haldane2023}. The tetratic order parameter is a symmetric traceless rank-$4$ tensor $Q_{abcd}$.
Importantly, in two space dimensions, like for the nematic order parameter, the traceless symmetric tensor $Q_{abcd}=Q_T(8 d_a d_b d_c d_d - 8 \delta_{(ab}d_c d_{d)}+\delta_{(ab}\delta_{cd)})$,
where ${\bf{d}}= ( \cos \vartheta/4, \sin \vartheta/4)$ is the planar tetratic director, admits only two independent components that can be used to define the complex order parameter
\begin{equation} \label{tetraticQ}
\mathcal{Q}_T=Q_{1111}+i Q_{1112}= Q_T\, e^{i \vartheta} .
\end{equation}
This peculiar feature that holds only in two dimensions allows us to generalize most of the results derived in \cite{Maciejko} in the long-wavelength limit. 
In fact, also in the current case, we can define an action for the tetratic Laughlin states, which reads
\begin{eqnarray}\label{Landau-FQHE3}
\mathcal{L}_{TFQHE}= \mathcal{L}_{CS}+\mathcal{L}_T - \frac{1}{2 \pi}
\epsilon^{\mu\nu\lambda} a^T_{\mu} \partial_{\nu} a_{\lambda}, \hspace{0.7cm} \nonumber \\
a^T_c = - \frac{1}{8} \epsilon_{bd} Q_{afgb} \partial_c Q_{afgd}+ G \epsilon_{bd}\epsilon_{fg} \epsilon_{mn}\partial_b \partial_f \partial_m Q_{dgnc}, \nonumber \\ \hspace{0.3cm}
a^T_0 = - \frac{1}{8} \epsilon_{bd} Q_{afgb} \dot{Q}_{afgd}, \hspace{1.3cm}
\end{eqnarray}
where $G$ is a constant parameter.
Close to the transition from an isotropic to a tetratic FQH phase, the corresponding effective LdG Lagrangian $\mathcal{L}_T$ \cite{Giomi2, Giomi3} augmented by the Berry phase term $a^T_0$ is given by
\begin{equation}\label{Landau3}
\mathcal{L}_{T} = \frac{k'_1}{8} \epsilon_{bd} Q_{afgb} \dot{Q}_{afgd} - \frac{k'_2}{8} (\partial_f Q_{abcd})^2  -\frac{k'_3}{2} (Q_{abcd})^2,
\end{equation}
where $k'_i$ are some given physical parameters.
Similarly to the nematic case, we have
\begin{eqnarray}\label{Hamiltonian5}
H_{T}= \frac{k'_2}{8} |\partial_i \mathcal{Q}_T|^2+\frac{k'_3}{2} |\mathcal{Q}_T|^2,
\end{eqnarray}
that can be diagonalized in momentum space
\begin{align}\label{Hamiltonian6}
\mathcal{H}_{T}= \int \frac{d^2 k}{(2 \pi)^2} E_{\bf{k}} \gamma_{\bf{k}}^\dagger \gamma_{\bf{k}}, \hspace{0.06cm} [\gamma_{\bf{k}}, \gamma^\dagger_{\bf{k'}}] = (2 \pi)^2 \delta({\bf{k-k'}}),
\end{align}
where $\gamma_{\bf{k}}^\dagger$ are bosonic creation operators and
$E_{\bf{k}}= m' + ({k'_2}/4){\bf{k}}^2$,
where $m'= k'_3$ is a constant mass term. Thus, the mode spectrum identifies a gapped spin-$4$ collective mode.
All the results presented above can be naturally generalized to spin-p modes associated to $C_\textrm{p}$ symmetries and p-atic Hall states.
See also \cite{Cappelli} for a complementary way to introduce effective bulk higher-spin fields in the FQHE, that is however motivated by area-preserving diffeomorphisms rather than by p-atic phases.

\section{From Fierz-Pauli equations to higher-spin Schroedinger actions using symmetries}

The LdG effective-field-theory description of the GMP mode in \eqref{Landau} was first proposed in \cite{Bergshoeff:2017vjg}. We now re-derive it by concentrating on symmetries only, before using the same approach to derive the spin-4 LdG effective theory \eqref{Landau3} and discuss how to encompass also massive modes of higher spin. 

\subsection{Nematic order parameter}

Our starting point are the equations of motion for a free relativistic spin-$2$ particle of mass $m$:
\begin{equation}\label{FP}
\left(\Box -m^2 c^2\right) h_{\mu\nu}=0,\qquad\partial_\mu h^{\mu\nu}=0= \eta_{\mu\nu} h^{\mu\nu},
\end{equation}
where $h_{\mu\nu} = h_{\nu\mu}$ is a symmetric tensor field and
$\eta^{\mu\nu} = \textrm{diag} (-1/c^2, 1,1).$
The Fierz-Pauli (FP) equations \eqref{FP} describe two modes  with helicities $\pm 2$ and are invariant under translations and Lorentz transformations.
We decompose the traceless field $h_{\mu\nu}$ as follows
\begin{equation}\label{parspin-2}
h_{\mu\nu} =
\left(
\begin{array}{c|c}
 \phi & v_b\\ 
 \hline
v_a&\frac{1}{2c^2} \phi\, \delta_{ab} +S_{ab}
\end{array}
\right),\hskip .5truecm a=1,2
\end{equation}
with $S_{ab}$ symmetric and traceless. 
We then observe that under a Lorentz boost a time index transforms to a spatial index but that the inverse transformation from a spatial index to a time index is suppressed by a factor of $1/c^2$ (see Eq.~\eqref{spin-2_Lorentz}).
This implies that, after taking the limit $c \to \infty$, the boost becomes a transformation that only acts on the different components of $h_{\mu\nu}$ as follows
\begin{equation} \label{chain_2}
\phi \longrightarrow v_a\longrightarrow S_{ab}\longrightarrow 0 .
\end{equation}
From this we deduce that, unlike the $\phi$ and $v_a$ components, the symmetric traceless field $S_{ab}$ forms by itself a representation of the Galilei group, which is obtained as a $c \to \infty$ contraction of the Poincar\'e group.
In $2+1$ dimensions one can combine the two independent components $S_{11}$ and $ S_{12}$ of $S_{ab}$ into the complex field $\mathcal{S}_N =S_{11} + i S_{12}$, that transforms under infinitesimal spatial rotations generated by $\Lambda_{ab} = \epsilon_{ab}\lambda$  not as a scalar but as
\begin{equation}
\mathcal{S}_N^\prime(x^\prime) = \mathcal{S}_N(x)-2i\lambda \mathcal{S}_N(x).
\end{equation}
Furthermore, we find that $\mathcal{S}(x)$ transforms under internal  parity (P) $x^1 \to -x^1$, time-reversal (T) $t \to -t$  and inversion (I) $x^{1,2} \to - x^{1,2}$ as follows
\begin{align}\label{discrete}
\textrm{P:}\hskip .2truecm \mathcal{S}_N^\prime(x^\prime) = \mathcal{S}_N^\star(x)\,,\hskip 0.2truecm 
\textrm{T, I:}\hskip .2truecm
 \mathcal{S}_N^\prime(x^\prime) = \mathcal{S}_N(x).
\end{align}

It is well known that the Galilei group cannot be used to describe a massive mode. One argument is that at the non-relativistic level one expects two Noether symmetries describing the conservation of energy and mass. 
The second Noether symmetry extends the Galilei algebra to the Bargmann algebra \cite{Bargmann:1954gh}. 
To achieve this extension we redefine, before taking the limit $c \to \infty$, the complex field $\mathcal{S}_N$ in terms of another complex field $\mathcal{Q}_N$ as follows \cite{Bergshoeff:2017vjg}
\begin{equation}\label{Bargmann}
\mathcal{S}_N = e^{-i(mc^2 - E_0)t}\mathcal{Q}_N,
\end{equation}
thereby introducing a new constant $E_0$. 
The phase factor has the effect that after taking the limit $c \to \infty$, the field $\mathcal{Q}_N(x)$ transforms under boosts with parameter $\Lambda_a$ as
\begin{equation}\label{transfB}
\delta \mathcal{Q}_N = -(\Lambda^a t)\partial_a \mathcal{Q}_N + im\Lambda_a x^a \mathcal{Q}_N,
\end{equation}
where the last new term leads to a Bargmann
U(1) central extension. Furthermore, in the limit $c \to \infty$ the symmetry under parity P and time-reversal T is broken. This is due to the fact that both the $t\to -t$ time-reversal transformation and the complex conjugation that occurs in the parity transformation \eqref{discrete} leaves a factor $e^{-imc^2}$ in the transformation rule of the non-relativistic field $\mathcal{Q}_N$ which does not have a well defined $c \to \infty$ limit.  Instead, we find that the inversion symmetry I is preserved.
There exists a unique action invariant under translations, spatial rotations together with the modified boost of Eq.~\eqref{transfB}, that is of first order in the time derivatives, thus breaking the invariance under time-reversal symmetry. It is given by the spin-$2$ planar Schroedinger action
\begin{eqnarray}\label{Schroedinger}
I_{\rm Schroedinger} \propto  \int d^3 x \,  &&\Big [
 im\mathcal{Q}_N  \mathcal{\dot Q}_N^*
 -i m \mathcal{Q}_N^* {\mathcal{\dot Q}}_N + 
 \nonumber\\[.1truecm]
 && \hspace{-0.5cm}\partial^a\mathcal{Q}_N^* \partial_a\mathcal{Q}_N + 2mE_0\mathcal{Q}_N\mathcal{Q}_N^*\Big].
 \end{eqnarray}
This action corresponds to the complex rewriting of Eq.~\eqref{Landau}, thus allowing one to interpret $\mQ_N$ as the nematic order parameter \eqref{nematic}.
Whereas the FP equations \eqref{FP} describe two helicity modes, that under parity are mapped to each other, the Schroedinger action \eqref{Schroedinger} describes just a single mode and parity is broken. In the following section we will also derive the Schroedinger action \eqref{Schroedinger} from a direct $c \to \infty$ limit of a relativistic action, thus showing how the second helicity mode drops out in the non-relativistic limit, consistently with the previous discussion.
In summary, we conclude that the LdG description of the fluctuations of the nematic phase order parameter around the isotropic phase in the long wavelength limit given in Eq.~\eqref{Landau} can be recovered from a non-relativistic limit of  the three-dimensional FP theory. 

\subsection{Tetratic order parameter}

We can now generalize the above discussion to the spin-$4$ case, which is the next higher-spin field that preserves inversion symmetry.  
We start from a rank-$4$ symmetric tensor field $\Phi_{\mu\nu\rho\sigma}(x)$ that satisfies the spin-4 FP equations \cite{Fierz:1939ix}
\begin{equation}
\left(\Box -m^2 c^2\right) \Phi_{\mu\nu\rho\sigma}=0, \quad
\partial_\mu \Phi^{\mu\nu\rho\sigma}=0= \eta_{\mu\nu} \Phi^{\mu\nu\rho\sigma}.
\end{equation}
This field can be decomposed into the different helicity modes as $\Phi_{\mu\nu\rho\sigma} = \{ \beta\,, \alpha_a\,, \phi_{ab}\,, v_{abc}\,, S_{abcd}\}$, where each component is completely symmetric and traceless in its spatial indices, so that it describes two helicity modes except for $\beta$.
The independent components of $\Phi_{\mu\nu\rho\sigma}$ decompose in analogy with \eqref{parspin-2} as
\beq\label{dec-spin4}
\bal
\Phi_{abcd}&=S_{abcd}+\frac1{6c^2} \phi_{(ab}\delta_{cd)}+\frac1{8c^4} \delta_{(ab}\delta_{cd)}\beta , \\
\Phi_{abc0}&=v_{abc}+\frac1{4c^2} \alpha_{(a}\delta_{bc)} ,
\eal
\eeq
with parentheses denoting a symmetrization, where dividing by the number of terms in the symmetrization is understood. The decomposition of the remaining components follows from the traceless condition $\eta_{\mu\nu} \Phi^{\mu\nu\rho\sigma} = 0$. Under boost transformations these components transform, after taking the limit  $c \to \infty$, as follows
\begin{equation} \label{chain_4}
\beta  \longrightarrow \alpha_a\longrightarrow \phi_{ab}\longrightarrow v_{abc} \longrightarrow S_{abcd}\longrightarrow 0.
\end{equation}
This shows that the highest component $S_{abcd}$ forms a representation of the Galilei algebra by itself. Defining 
$\mathcal{S}_T = S_{1111} + i S_{1112}$,
we find that under spatial rotations the complex field $\mathcal{S}$ transforms as 
\begin{equation}
\mathcal{S}_T^\prime(x^\prime) = \mathcal{S}_T(x)-4i\lambda \mathcal{S}_T(x),
\end{equation}
which shows that we are dealing with a spin-$4$ field. 
Making the same redefinition \eqref{Bargmann} as in the spin-$2$ case,
\begin{equation}\label{Bargmann2}
\mathcal{S}_T = e^{-i(mc^2 - E_0)t}\mathcal{Q}_T,
\end{equation}
we find in the $c\to \infty$ limit the same continuous and discrete symmetries as in the spin-$2$ case except for the spatial rotations.
Looking for an action invariant under these symmetries leads to a planar spin-$4$ Schroedinger action for $\mathcal Q_T$, 
which has the same form as the planar spin-$2$ Schroedinger action given in Eq.~\eqref{Schroedinger}. 
The latter in turn is equivalent to the tetratic LdG Lagrangian \eqref{Landau3} written in complex form. A similar analysis applies to the spin-$s$ FP equations \footnote{See also \cite{Horvathy:2010vm} and references therein for a group-theoretical discussion of the non-relativistic limit of higher-spin representations of the three-dimensional Poincar\'e group.}.

\section{From massive higher-spin actions to p-atic LdG actions}\label{sec:actions}

The LdG description for p-atic phases in the FQHE can also be recovered from the $c \to \infty$ limit of relativistic actions describing massive higher-spin fields. Let us stress that, aside from providing a complementary viewpoint on our previous derivation, moving to an off-shell derivation of the Schroedinger actions might provide a rationale for introducing non-linearities and background couplings as limits of their relativistic counterparts (see, e.g., \cite{Boulanger:2018dau, Zinoviev:2022fmz} for recent works on massive higher-spin couplings), although complementary on-shell techniques have been explored too \cite{Bergshoeff:2011pm}. As in the previous section, we first show how to recover the spin-$2$ LdG action and then we generalize the analysis to the spin-$4$ case.

\subsection{Nematic order parameter}

We wish to rederive Eq.~\eqref{Schroedinger} starting from the non-relativistic limit of an action leading to the FP equations of motion \eqref{FP}. Motivated by the description of the spin-$2$ GMP mode in terms of unimodular gravity \cite{Gromov}, we choose to work with the action resulting from the coupling of linearized unimodular gravity with auxiliary Stueckelberg fields $B_\mu$ and $S$,
\beq\label{lag_spin-2}
\begin{split}
&\kappa^{-1} \cL =\frac12 h^{\mu\nu} \left(  \Box h_{\mu\nu} - 2\pr_\mu \prdt h_\nu  -m^2c^2 h_{\mu\nu}
\right)  \\
&+ \frac12 B^{\mu} \left( \Box B_{\mu} - \pr_\mu \prdt B 
\right)  
+\frac12 S \left(\Box S +3 m^2 c^2 S\right)
\\
&+ \sqrt2 mc\, h^{\mu\nu}\pr_{\mu} B_{\nu}
+ 2mc\, B^\mu \pr_\mu S ,
\end{split}
\eeq
where $h_{\mu\nu}$ is traceless and $\kappa$ is the gravitational constant, as we are assuming that this action results from the linearization of a massive gravity theory. This strategy to build quadratic actions for massive fields is the analogue of that described, e.g., in \cite{Zinoviev:2001dt, Zinoviev:2008ze} and we provide more details in App.~\ref{app:proca}.
This action is invariant under the gauge transformations
\begin{equation}\label{gauge2}
\begin{split}
&\delta h_{\mu\nu}  = \pr_{\mu} \epsilon_{\nu} + \pr_{\nu} \epsilon_{\mu} - \frac{2}{3}\, \eta_{\mu\nu} \prdt \epsilon , \\
&\delta B_\mu  = \pr_\mu \lambda + \sqrt{2} m c\, \epsilon_\mu ,\quad
\delta S  = 2m c\,\lambda,
\end{split}
\end{equation}
where the gauge parameters satisfy the constraint 
\begin{equation}
\prdt \epsilon -\frac{3mc}{\sqrt2}\,\lambda = 0.
\end{equation}
As discussed in App.~\ref{app:proca}, the equations of motion derived from the action \eqref{lag_spin-2} reduce to the FP ones, i.e., to \eqref{FP}.
We can also gauge away the scalar field $S$ using its algebraic gauge transformation. This leads to a transverse condition $\prdt \epsilon=0$ for the vector parameter, implying that only residual volume-preserving diffeomorphisms in $2+1$ dimensions are allowed. 

We now decompose $h_{\mu\nu}$ as in \eqref{parspin-2} in terms of the spatial tensors $S_{ab}$, $v_a$ and $\phi$ and we introduce $B_\mu=(v,S_a)$. 
We also introduce the complex fields 
\beq \label{def_S}
\begin{split}
\mS_{[2]}& = \mS_N = S_{11}+iS_{12} , \quad
\mS_{[1]}= S_1+iS_2 , \\
\mV_{[1]}& =v_1+iv_2.
\end{split}
\eeq
We then define
\begin{equation} \label{V-U}
\mS_{[k]}=e^{-i(mc^2-E_0)t}\mQ_{[k]}, \hspace{0.3cm}
\mV_{[k]}=e^{-i(mc^2-E_0)t}\mU_{[k]},
\end{equation}
where $k$ denotes the rank of the field, while $m$ coincides with the mass parameter in the action \eqref{lag_spin-2} and $E_0$ is an arbitrary constant. The resulting complex field $\mQ_{[2]}$ has to be identified with the $\mQ_{N}$ introduced in Eq.~\eqref{nematic}. 
Integrating out the field $\mathcal U_{[1]}$ and taking the limit $c \to \infty$ leads to 
\beq\label{SchLagrangianspin2}
\kappa^{-1}\mathcal L= 2i m\dot \mQ_{[2]} \mQ_{[2]}^*-2 mE_0 \mQ_{[2]} \mQ_{[2]}^* 
-\partial_a \mQ_{[2]} \partial_a \mQ_{[2]}^* \, ,
\eeq
where we omitted all terms that do not depend on $\mQ_{[2]}$ and thus decouple (see App.~\ref{app:limit}). 
In the non-relativistic limit we thus recover the nematic LdG Lagriangian \eqref{Landau} with parameters: $k_1= -8\kappa m$, $k_2= 4\kappa$, $k_3= 2\kappa m E_0$.

\subsection{Tetratic order parameter}

We now turn our attention to a relativistic massive spin-$4$ field \footnote{A non-relativistic limit similar to that discussed below was considered in \cite{Dutta:2023cpc} employing, however, a different action and a different rescaling of the fields.}. The analogue of the Stueckelberg Lagrangian \eqref{lag_spin-2} involves the Maxwell-like kinetic terms that have been introduced in \cite{Skvortsov:2007kz, Campoleoni:2012th} to generalize linearized unimodular gravity and reads
\beq \label{relativistic-spin-4}
\bal
\mathcal L& = \sum_{k=0}^4 \frac{\kappa}{2}\, \varphi^{\mu(k)} \bigg[ \left(\Box -\frac{(k-3)(k+5)m^2c^2}{2k+1}\right) \varphi_{\mu(k)} \\
& - k\, \pr_\mu \prdt \varphi_{\mu(k-1)}  + 2mc\sqrt{\frac{k(5-k)(k+3)}{2k-1}}\, \pr_\mu \varphi_{\mu(k-1)} \bigg] ,
\eal
\eeq
where $\kappa$ is the gravitational constant, $\mu(k)$ stands for a set of $k$ symmetrized indices, while repeated indices denote a symmetrization. The coefficients in Eq.~\eqref{relativistic-spin-4} are fixed by demanding the invariance of the action under a deformation of the gauge symmetry of each kinetic term (see App.~\ref{app:proca}).
We then decompose the rank-$4$ field $\varphi_{\mu(4)}\equiv \Phi_{\mu\nu\rho\sigma}$ in traceless spatial components
as in \eqref{dec-spin4} and we introduce the complex variables $\mS_{[4]} = \mS_T =S_{1111}+iS_{1112}$ and $\mV_{[3]}=v_{111}+iv_{112}$. 
We also introduce a similar decomposition for the auxiliary fields of rank $k < 4$.
After redefining the fields as in Eq.~\eqref{V-U} and integrating out from the action the field $\mU_{[3]}$, the field $\mQ_{[4]}$ (which coincides with $\mathcal{Q}_T$ in Eq.~\eqref{tetraticQ}) decouples from the other fields in the limit $c\rightarrow \infty$ (see App.~\ref{app:proca} for more details). 
Its dynamics is then described by the Schroedinger Lagrangian in Eq.~\eqref{Landau3}, with the identifications $k'_1= -8\kappa m$, $k'_2= 4\kappa$, $k'_3= 2\kappa m E_0$. 

\section{Measurement proposal for the spin-4 mode}

Recently, an inelastic light scattering experiment using circularly polarized photons has measured the chirality of the GMP mode in Gallium Arsenide (GaAs) quantum wells \cite{Pinczuk}, confirming its geometric interpretation as a massive graviton-like mode. Here, we propose a generalization of the experimental approach in \cite{Pinczuk}, to detect the spin-4 mode.
According to angular momentum conservation, the angular momentum transferred to a FQH liquid equals the change in photon spin during light scattering. For instance, a GMP mode with spin $s=-2$ is excited when the incident photon has spin $-1$ and the scattered photon has spin $+1$. Formally, to detect a collective higher-spin mode with spin
$s=-4$, by adopting a similar approach, we would need to employ an incident polarized graviton with spin $-2$ and measure a scattered polarized graviton with spin $+2$.
Clearly, there are no current experiments involving real gravitons. However, effective gravitons could be emulated with light by following the method presented in \cite{Cirio}, in which it has been theoretically shown that a circularly polarized gravitational wave can be emulated by two circularly polarized electromagnetic waves with parallel momenta.
In other words, two circularly polarized photons with the same helicity and parallel momenta will be required in an inelastic light scattering experiment to detect the spin-4 mode in FQH states \footnote{We wish to thank Lingjie Du for inspiring discussions about this proposal}.

\section{Conclusions and outlook}

We have shown and discussed the relation between p-atic phases and non-relativistic higher-spin fields in the FQHE. 
We have derived the effective field theories for the higher-spin collective modes in the Laughlin states, given by planar spin-p Schroedinger actions, from the non-relativistic limit of $(2+1)$-dimensional relativistic massive higher-spin theories. 
There are several important open questions that we aim to investigate in future work: i) the effective-field-theory description of higher-spin collective modes in the presence of a curved background \cite{Gromov, Klevtsov2014, Klevtsov2015, Can, Salgado-Rebolledo1}. Indeed, by coupling the higher-spin fields to a background geometry, we expect to obtain novel geometric terms beyond the Wen-Zee term. Once the higher-spin fields are integrated out, the resulting effective action will lead, in addition to the Wen-Zee term induced by the spin-2 mode \cite{Maciejko,Gromov3}, to extra terms similar to those derived in Ref.~\cite{Cappelli}. Since these new terms uniquely depend on the probing fields and the background geometry, they can, in principle, identify measurable quantities. 
Moreover, it would be interesting to see whether one can construct a planar spin-p Schroedinger action in a Newton-Cartan background, generalizing the spin-$0$ analysis of \cite{Bergshoeff:2015sic};
ii) a supergeometric extension to describe the spin-$3/2$ massive mode \cite{Salgado-Rebolledo2, Gromov4, Pu2023} and possible further semi-integer higher-spin modes;
iii) higher-spin interactions allowing to describe bulk modes away from the isotropic point. 
For all these developments, one could also exploit the $c \to \infty$ limit of the plethora of relativistic higher-spin interactions that only exist in three dimensions. 
These have been mainly studied for massless higher-spin fields (see \cite{Campoleoni:2024ced} for a review), but investigations of massive interactions appeared, e.g., in \cite{Chen:2011yx, Zinoviev:2022fmz};
iv) more general higher-spin field theories possibly describing incompressibility and collective modes in higher-dimensional FQH fluids \cite{Palumbo7}.
Finally, we wish to mention that in three dimensions there exist alternative descriptions of the dynamics of massive helicities in terms of higher-derivative theories, for both the spin-$2$ and higher-spin cases; see, e.g., \cite{Damour:1987vm, Bergshoeff:2009fj, Bagchi:2011vr, Chen:2011yx, Bergshoeff:2011pm, Dalmazi:2021dgp} and references therein. We defer to future work an analysis of our non-relativistic limit within this approach.

\vspace{0.1cm}
\noindent {\bf Acknowledgements:} A.C.\ is a research associate of the Fonds de la Recherche Scientifique -- FNRS. His work was supported by FNRS under Grants F.4503.20 and T.0047.24. P.S.-R.\ was supported by the Austrian Science Fund (FWF), projects P 33789 and P 36619, and by the Norwegian Financial Mechanism 2014-2021 via the Narodowe Centrum Nauki (NCN) POLS grant 2020/37/K/ST3/03390. We thank the Erwin Schr\"odinger International Institute for Mathematics and Physics and P.S.-R.\ thanks the University of Mons for hospitality. We are pleased to acknowledge discussions with D.~Francia, Z.~Papi\ifmmode \acute{c}\else \'{c}\fi{} and S.~Pu,  L. Du and B. Yang.

\appendix

\section{Transformations of the relativistic fields under Lorentz boosts}\label{app:lorentz}

We recall here the Poincar\'e transformations of the relativistic fields of spin two and four on which we built our symmetry driven derivation of the Schroedinger actions in Eqs.~\eqref{Landau} and \eqref{Landau3}.

We begin from the spin-$2$ field $h_{\mu\nu}$, which transforms under translations and Lorentz transformations with
infinitesimal parameters $a^\mu$ and $\Lambda_{\mu\nu} = -\Lambda_{\nu\mu}$ as
\begin{equation}\label{symmetries}
h_{\mu\nu}^\prime (x^\prime) = \frac{\partial x^\rho}{\partial x^{\prime \mu}}\frac{\partial x^\sigma}{\partial x^{\prime \nu}}\, h_{\rho\sigma}(x)
\end{equation}
with
\begin{equation}\label{parameters}
x^{\prime \mu} = x^\mu - \xi^\mu(x)\hskip .5truecm \textrm{and}\hskip .5truecm \xi^\mu(x) = a^\mu -\Lambda^\mu{}_\nu\,x^\nu.
\end{equation}
This leads to the infinitesimal variations
\beq
\delta h_{\mu\nu}(x) = h'_{\mu\nu}(x) - h_{\mu\nu}(x) = \xi^\lambda \partial_\lambda h_{\mu\nu} + 2\partial_{(\mu} \xi^\rho h_{\nu)\rho},
\eeq
where $(\mu_1\cdots \mu_s)$ denotes a normalized symmetrization of the enclosed indices, e.g.,  $A_{(\mu} B_{\nu)} := \frac{1}{2} \left( A_{\mu} B_{\nu} + A_{\nu} B_{\mu} \right)$. In terms of the spatial traceless components introduced in Eq.~\eqref{parspin-2}, these transformations read
\begin{subequations} \label{spin-2_Lorentz}
\begin{align}
\delta \phi & = \xi^\lambda \partial_\lambda \phi - 2 \Lambda^a v_a , \\
\delta v_a & = \xi^\lambda \partial_\lambda v_a + \Lambda_a{}^b v_b - \Lambda^b S_{ab} + O(c^{-2}) , \\
\delta S_{ab} & = \xi^\lambda \partial_\lambda S_{ab} + 2\Lambda_{(a}{}^c S_{b)c} + O(c^{-2}) ,
\end{align}
\end{subequations}
where $a,b \in \{1,2\}$ and we denoted the generator of Lorentz boosts as $\Lambda_{a} := \Lambda_{a0}$, while $\Lambda_{ab}$ is the generator of spatial rotations.
These variations show that in the limit $c \to \infty$ the chain of transformations in Eq.~\eqref{chain_2} holds. In particular, this implies that the symmetric traceless field $S_{ab}$ forms by itself a representation of the Galilei group, which is obtained as a $c \to \infty$ contraction of the Poincar\'e group as mentioned in the main text.

A spin-$4$ field transforms instead under translations and Lorentz transformations generated by the same infinitesimal parameters as
\beq
h_{\mu\nu \rho\sigma}^\prime (x^\prime) = \frac{\partial x^\alpha}{\partial x^{\prime \mu}}\frac{\partial x^\beta}{\partial x^{\prime \nu}}
 \frac{\partial x^\gamma}{\partial x^{\prime \rho}}\frac{\partial x^\delta}{\partial x^{\prime \sigma}}\, h_{\mu\nu\rho\sigma}(x) ,
\eeq
with $x'^\mu$ given again by Eq.~\eqref{parameters}. This implies the infinitesimal variations
\beq
\delta h_{\mu\nu\rho\sigma} = \xi^\lambda\partial_\lambda h_{\mu\nu\rho\sigma} + 4\partial_{(\mu}\xi^\lambda h_{\nu\rho\sigma)\lambda} ,
\eeq
that in terms of the traceless components introduced in Eq.~\eqref{dec-spin4} read
\begin{align}
\delta \beta & = \xi^\lambda \partial_\lambda \beta - 4 \Lambda^a \alpha_a , \\
\delta \alpha_a & = \xi^\lambda \partial_\lambda \alpha_a + \Lambda_{a}{}^b \alpha_b - 3 \Lambda^b \phi_{ab} + O(c^{-2}) , \\
\delta \phi_{ab} & = \xi^\lambda \partial_\lambda \phi_{ab} + 2 \Lambda_{(a}{}^c \phi_{b)c} - 2 \Lambda^c v_{abc} + O(c^{-2}) , \\
\delta v_{abc} & = \xi^\lambda \partial_\lambda v_{abc} + 3 \Lambda_{(a}{}^d \phi_{bc)d} - \Lambda^d S_{abcd} + O(c^{-2}) , \\
\delta S_{abcd} & = \xi^\lambda \partial_\lambda S_{abcd} + 4 \Lambda_{(a}{}^e S_{bcd)e} + O(c^{-2}) ,
\end{align}
corresponding to the chain of transformations summarized in Eq.~\eqref{chain_4}. 

\section{Proca-like actions for massive particles of any spin} \label{app:proca}

In Sect.~\ref{sec:actions}, we presented quadratic actions describing relativistic massive spin-$s$ fields that involve a number of auxiliary fields. The rationale for building these actions is that a massive spin-$s$ field in $D$ spacetime dimensions propagates the same helicities as a tower of massless fields of spin comprised between $0$ and $s$, with each spin appearing once. As a result, one can build a quadratic action for a massive spin-$s$ field by coupling the free actions for massless fields with spin comprised between $0$ and $s$ while preserving a deformation of their free gauge symmetry.

To fix the ideas, let us consider the case of Proca's theory, describing a massive spin-$1$ field. The Proca Lagrangian
\begin{equation} \label{proca}
\cL = - \frac{1}{4} \, F^{\mu\nu} F_{\mu\nu} - \frac{m^2c^2}{2}\, B_\mu B^\mu ,
\end{equation}
with $F_{\mu\nu}=\partial_\mu B_\nu-\partial_\nu B_\mu$, does not have any gauge symmetry. One can however reinstate the gauge symmetry of the Maxwell Lagrangian by introducing a ``Stueckelberg field'' via the shift
\begin{equation}
B_\mu \to B_\mu - \frac{1}{m c}\,\pr_\mu S .
\end{equation}
The resulting Lagrangian reads
\begin{equation} \label{proca-stueck}
\cL = - \frac{1}{4} \, F^{\mu\nu} F_{\mu\nu} - \frac{m^2c^2}{2}\, B_\mu B^\mu - \frac{1}{2}\, \pr_\mu S\, \pr^\mu S + mc\, B^\mu \pr_\mu S,
\end{equation}
and it is invariant under the gauge transformations
\begin{equation} \label{transf-proca}
\delta B_\mu = \pr_\mu \lambda , \qquad
\delta S = mc\, \lambda .
\end{equation}
The gauge transformation of the scalar field is algebraic: this means that one can use it to eliminate $S$ and go back to the Lagrangian \eqref{proca}. We denote all fields with an algebraic gauge transformation as Stueckelberg fields.
We obtained the Lagrangian in Eq.~\eqref{proca-stueck} performing a shift in the Proca Lagrangian. Alternatively, one can also consider the most general coupling of a Maxwell field with a free scalar and fix the coefficients demanding that the action be invariant under the transformations \eqref{transf-proca}. 

This strategy was used in Ref.~\cite{Zinoviev:2001dt} to describe a massive spin-$s$ field via the coupling of $s$ Fronsdal's Lagrangians \cite{Fronsdal:1978rb}, describing massless fields of spin $0 \leq k \leq s$ (see also Ref.~\cite{Rindani:1985pi}). Fixing completely the algebraic Stueckelberg symmetries then gives the Singh-Hagen Lagrangian \cite{Singh:1974qz}, which describes a massive spin-$s$ field in terms of a traceless rank-$s$ tensor and a number of auxiliary fields growing with $s$.

In this paper, we used a similar strategy, but substituting Fronsdal's Lagrangians with Maxwell-like ones \cite{Skvortsov:2007kz, Campoleoni:2012th}. 
Within this framework, the Lagrangian describing a massive spin-$s$ field contains $s$ traceless fields that we distinguish by the number of indices they carry: $\{\varphi_{\mu(k)}\}$ with $0 \leq k \leq s$. We recall that the shorthand $\mu(k)$ denotes a set of $k$ symmetrized indices; for instance, $\varphi_{\mu(2)} := \varphi_{\mu_1\mu_2}$. The Lagrangian reads
\begin{equation} \label{Ls}
\begin{split}
\cL = \frac{\kappa}{2} \sum_{k=0}^s  \varphi^{\mu(k)} \Big( \Box \varphi_{\mu(k)} - k\, \pr_\mu \pr\cdot \varphi_{\mu(k-1)} \\
- m_k\, (mc)^2\, \varphi_{\mu(k)} + 2 c_k\,mc\, \pr_\mu \varphi_{\mu(k-1)} \Big) , 
\end{split}
\end{equation}
with
\begin{subequations} \label{coeff-s}
\begin{align}
m_k & =- \frac{(s-k-1)(D+s+k-2)}{D+2k-2} , \label{mk} \\
c_k & = \sqrt{\frac{k(s-k+1)(D+s+k-4)}{D+2k-4}} . \label{ck}
\end{align}
\end{subequations}
Here and in the following, repeated covariant or contravariant indices denote a symmetrization, where dividing by the number of terms in the symmetrization is understood. For instance, $\pr_\mu \xi_\mu := \frac{1}{2} \left( \pr_{\mu_1} \xi_{\mu_2} + \pr_{\mu_2} \xi_{\mu_1} \right)$.
The action involving \eqref{Ls} is invariant under the gauge transformations
\begin{equation} \label{var-s}
\begin{split}
\delta \varphi_{\mu(k)} & = k\, \pr_\mu \xi_{\mu(k-1)} + c_{k+1}\,mc\, \xi_{\mu(k)} \\
& - \frac{2}{D+2k-4} \binom{k}{2} \eta_{\mu\mu} \pr\cdot \xi_{\mu(k-2)}  ,
\end{split}
\end{equation}
where the gauge parameters $\{ \xi_{\mu(k)}\}$, with $0 \leq k \leq s-1$, are traceless and satisfy the constraints
\begin{equation}
\pr\cdot \xi_{\mu(k-1)} - \frac{(D+2k-2)c_{k+1}}{(k+1)(D+2k-4)} \,mc\, \xi_{\mu(k-1)} = 0 , 
\end{equation}
with $1 \leq k \leq s-1$.
The Lagrangian \eqref{Ls} depends on a single free mass parameter $m$ and its coefficients can be fixed by demanding invariance under gauge transformations of the form \eqref{var-s}. In the limit $m \to 0$ one obtains the sum of $s$ Maxwell-like Lagrangians, which are invariant under gauge transformations generated by traceless and divergenceless gauge parameters, thus presenting themselves as the natural higher-spin generalization of unimodular gravity (linearized volume-preserving diffeomorphisms are indeed generated by transverse vectors).

Alternatively, the coefficients in the Lagrangian \eqref{Ls} can also be fixed by demanding that the equations of motion reduce to the Fierz-Pauli ones. Let us discuss this in the example of a massive spin-$2$ field. To this end, it is convenient to first eliminate the scalar field by using its algebraic Stueckelberg symmetry. After this step, the Euler-Lagrange equations of \eqref{Ls} read
\begin{subequations} \label{eom2}
\begin{align}
& \Box h_{\mu\nu} - 2\,\pr_{(\mu} \pr\cdot h_{\nu)} - \frac{2}{D}\, \eta_{\mu\nu} \pr\cdot\pr\cdot h - m^2c^2 h_{\mu\nu} \nonumber \\
& \quad + c_2\, mc \left( \pr_{(\mu} B_{\nu)} - \frac{1}{D}\, \eta_{\mu\nu} \pr\cdot B \right) = 0 \, , \label{eom-h} \\[5pt]
& \Box B_\mu - \pr_\mu \pr\cdot B - c_2\,mc\, \pr\cdot h_\mu = 0 \, , \label{eom-A}
\end{align}
\end{subequations}
where we renamed the fields as in Eq.~\eqref{lag_spin-2} to distinguish them more easily. Notice that the equations of motion must be traceless as the fields entering the Lagrangian. For instance, the double divergence of $h_{\mu\nu}$ appears in Eq.~\eqref{eom-h} as a result of the traceless projection. Taking a divergence of Eq.~\eqref{eom-A} one obtains $\pr\cdot\pr\cdot h = 0$ and substituting this result in the divergence of Eq.~\eqref{eom-h} one obtains
\begin{equation}
\frac{(D-1)}{D}\,c_2\, \pr_\mu \pr\cdot B = \left( 1 - \frac{c_2^2}{2} \right) m c\, \pr\cdot h \, .
\end{equation}
The value of $c_2$ fixed by Eq.~\eqref{ck} is precisely that setting to zero the right-hand side. This allows one to conclude $\pr_\mu \pr\cdot B = 0$ and, eventually, $\pr\cdot B = 0$ up to constant functions that do not affect the propagated local degrees of freedom. This on-shell constraint is crucial because, after the gauge fixing setting the scalar to zero, the spin-$2$ gauge parameter satisfies $\pr\cdot \xi = 0$, so that the divergence of the vector potential becomes gauge invariant. On the other hand, the latter is forced to vanish by the equations of motion, so that one can gauge away the remaining portion of the vector potential by using the algebraic gauge transformation generated by $\xi_\mu$. After one reaches the condition $B_\mu = 0$ by combining the equations of motion and a gauge fixing, Eq.~\eqref{eom-h} reduces to the Fierz-Pauli equation in Eq.~\eqref{FP}.

\section{Non-relativistic limit of the Proca-like actions} \label{app:limit}

In this Appendix we detail the Galilean limit of the higher-spin actions of App.~\ref{app:proca} that we discussed in Sect.~\ref{sec:actions}. We restrict the analysis to $D=2+1$ dimensions and we consider explicitly the examples of fields of spin two and four, while commenting on how to extend the limiting procedure to fields of arbitrary spin. 

We split the indices as $\mu=\{0,a\}$ and we consider coordinates $x^\mu=\{x^0\equiv t, x^a\}$, so that the Minkowski metric and its inverse are given by $\eta_{\mu\nu}= {\rm diag}(-c^2,1,1)$ and $\eta^{\mu\nu}= {\rm diag}(-1/c^2,1,1)$, respectively.

\vspace{5pt}

\paragraph*{\bf{Spin-2 case:}}
We begin by considering the Lagrangian \eqref{lag_spin-2} for a massive spin-2 field, which correspond to the $s=2$ and $D = 3$ instance of \eqref{Ls}. The action is invariant under the gauge transformations \eqref{gauge2}.
Using the $\lambda$-symmetry to set to zero the scalar field $S$, we find the Lagrangian
\beq\label{fpactionred1}
\begin{split}
&\kappa^{-1} \cL =\frac12 h^{\mu\nu} \left(  \Box h_{\mu\nu} - 2\pr_\mu \prdt h_\nu  -m^2c^2 h_{\mu\nu}
\right)  \\
&+ \frac12 B^{\mu} \left( \Box B_{\mu} - \pr_\mu \prdt B 
\right)  
+ \sqrt2 mc\, h^{\mu\nu}\pr_{\mu} B_{\nu} .
\end{split}
\eeq

We can now split $h_{\mu\nu}$ in traceless spatial components as in Eq.~\eqref{parspin-2}, while also splitting the Stueckelberg field as $B_\mu = (v,S_a)$.
In $2+1$ dimensions we can then combine the two independent components of the previous traceless tensors of rank $k \geq 1$ into the complex fields introduced in Eq.~\eqref{def_S}.
This allows one to rewrite the Lagrangian \eqref{fpactionred1} (up to total derivatives) as
\begin{widetext}
\beq\label{Lagrangianallterms}
\bal
\kappa^{-1} \mathcal L&=
-\frac{1}{8c^6}\dot \phi^2
-\frac{1}{4c^4}|\partial \phi|^2 
+\frac{1}{2c^2}|\dot \mS_{[2]}|^2
+ \frac{1}{2c^2}
\bar\partial \mathcal S_{[2]} \bar\partial \phi
+\frac{ m}{\sqrt2 c^3} \phi \dot v 
-\frac{3m^2 }{8c^2}\phi^2
+\frac{m}{2\sqrt2 c}\phi \partial \mS^*_{[1]} 
- \frac{m^2 c^2}{2} |\mS_{[2]}|^2
\\&
+\frac{1}{4c^2} \left(
|\dot \mS_{[1]}|^2 -2 \dot \mS_{[1]} \bar\partial v
+|\partial v|^2\right) 
-\frac14|\partial \mS_{[1]}|^2
+ \frac1{16}\left(\partial \mS^*_{[1]} + \bar \partial \mS_{[1]} \right)^2 
- \frac{mc}{\sqrt2}
\bar \partial \mathcal S_{[2]} \mS_{[1]}^*
 + \kappa^{-1} \mathcal L_{\mV_{[1]}} +\rm{c.c.} ,
\eal
\eeq
\end{widetext}
where c.c. denotes the complex conjugate and we defined the complex derivatives $
\partial=\partial_1+i\partial_2$ and $
\bar\partial= \partial_1-i\partial_2$. Furthermore, we have isolated the terms that depend on $\mV_{[1]}$ in 
\begin{widetext}
\beq \label{LV}
\bal
\kappa^{-1} \mathcal L_{\mV_{[1]}}&=
 \frac{m^2}{2} |\mV_{[1]}|^2
+\frac{1}{2c^2}|\partial \mV_{[1]}|^2 
-\frac{1}{c^2}
\bar \partial \mathcal S_{[2]}\dot  \mV_{[1]}^*
+\frac{1}{2c^4}\dot \mV_{[1]} \bar\partial  \phi
-\frac{1}{8c^2} \left(\partial \mV^*_{[1]} 
+ \bar \partial \mV_{[1]} \right)^2
\!-
\frac{m}{\sqrt2c}  \left(
\mV_{[1]}^* \partial v
+\dot \mS_{[1]} \mV_{[1]}^*
\right) ,
\eal
\eeq
\end{widetext}
Before taking the non-relativistic limit, we introduce the complex fields $\mQ_{[2]}$, $\mQ_{[1]}$ and $\mU_{[1]}$ via the field redefinition \eqref{V-U}.
In terms of the new fields the Lagrangian takes the form
\begin{widetext}
\beq\label{LagrangianonlyQ2v3}
\bal
\kappa^{-1} \mathcal L&=
-mE_0 |\mQ_{[2]}|^2 
- im \mQ_{[2]} \dot\mQ_{[2]}^*
-\frac{mE_0}{2} |\mQ_{[1]}|^2
-\frac{im}{2} \mQ_{[1]} \dot \mQ_{[1]}^*
+\frac{m^2 c^2}{4}|\mQ_{[1]}|^2
- \frac{mc}{\sqrt2}
\bar \partial \mathcal Q_{[2]} \mQ_{[1]}^*
\\&
-
\frac14 |\partial \mQ_{[1]} |^2
+\frac{1}{16}\left(e^{i(mc^2-E_0)t}\partial \mQ^*_{[1]} +e^{-i(mc^2-E_0)t} \bar \partial \mQ_{[1]} \right)^2 + \kappa^{-1} \mathcal L_{\mU_{[1]}} +\rm{c.c.} +O(c^{-1}).
\eal
\eeq
\end{widetext}
Note that the field redefinition \eqref{V-U} introduced some factors $e^{-i(mc^2-E_0)t}$ that will oscillate in the $c \to \infty$ limit. On the other hand, these terms will decouple in the non-relativistic limit, so that we will ignore this issue as in \cite{Bergshoeff:2017vjg}. After introducing the new variables, the terms in Eq.~\eqref{LV} take the form
\beq
\mathcal L_{\mU_{[1]}}=
\frac{\kappa\, m^2}{2} |\mU_{[1]} |^2
- \kappa\,\mU_{[1]}^*\mathcal J_{[1]} +O(c^{-2}) ,
\eeq
where we have defined
\beq
\bal
\mathcal J_{[1]}&=
-\frac{i m^2 c}{\sqrt2} \mQ_{[1]} +im \bar\partial \mQ_{[2]} \\
& +\frac{m}{\sqrt2c}\left(
 \partial v +iE_0\mQ_{[1]}+ \dot\mQ_{[1]}
\right).
\eal
\eeq
To proceed, we note that for large values of $c$ the field equations for $\mU_{[1]}^*$ lead to $\mU_{[1]}=\frac{1}{m^2} \mathcal J_{[1]}+O(c^{-2})$, and therefore integrating out $\mU_{[1]}$ yields the following effective Lagrangian
\beq\label{LeffU1}
\bal
\mathcal L^{eff}_{\mU_{[1]}} & = - \frac12 | \partial \mQ_{[2]}|^2
 -\frac{m^2c^2}{4}|\mQ_{[1]}|^2
 +\frac{mc}{\sqrt 2} \mQ_{[1]} \partial \mQ_{[2]}^*
\\
&
+\frac{im}{2}\mQ_{[1]}\left(\bar \partial v +\dot\mQ_{[1]}^*- iE_0\mQ_{[1]}^*\right) 
+ O (c^{-1})
\eal
\eeq
When replacing this expression for $\mathcal L_{\mU_{[1]}}$  in the Lagrangian \eqref{LagrangianonlyQ2v3}, the terms multiplied by powers of $c$ that diverge in the limit $c \to \infty$  cancel, and $\mQ_{[2]}$ decouples from the other fields. Thus, in the limit $c \to \infty$, up to total derivatives, we find the spin-2 Schroedinger Lagrangian of Eq.~\eqref{Schroedinger},
\beq\label{LagrangianonlyQ2v2}
\mathcal L= 2i\kappa m \dot\mQ_{[2]} \mQ_{[2]}^*
-\kappa |\partial \mQ_{[2]}|^2
-2\kappa mE_0 |\mQ_{[2]}|^2
+\dots ,
\eeq
where the dots denote terms that do not depend on $\mQ_{[2]}$. Using the fact that up to total derivatives $\epsilon_{bc}Q_{ab}\dot Q_{ac}=-2i\dot\mQ_{[2]} \mQ^*_{[2]}$, the $\mQ_{[2]}$-dependent part of this Lagrangian can be put in the form of Eq.~\eqref{Landau}, with coefficients $k_1= -8\kappa m$, $k_2= 4\kappa$ and $k_3= 2\kappa m E_0$.

\vspace{5pt}

\paragraph*{\bf{Spin-4 case:}}
We begin by parametrizing the fields entering the Lagrangian \eqref{Ls} with $s = 4$ in terms of traceless spatial components. 
The full set of relativistic fields can be rewritten in terms of a set of six symmetric and traceless tensors $\{S_{abcd},S_{abc},S_{ab},v_{abc},v_{ab},\phi_{ab}\}$, four vectors $\{S_a,v_a,\phi_a,\alpha_a\}$ and five scalar fields $\{\beta,\alpha,v,\phi,S\}$ as follows.
For the auxiliary fields of rank one and two we use the same notation as in our analysis of the spin-2 case and we similarly decompose the rank-three auxiliary field in traceless spatial components. We thus define
\beq\label{parchi}
\varphi_{ab0}=\frac{1}{2c^2}\alpha \delta_{ab}+v_{ab},\quad
\varphi_{abc}= \frac{3}{4c^2} \phi_{(a}\delta_{bc)}+S_{abc},
\eeq
while the remaining components are fixed by the constraint $\eta^{\mu\nu} \varphi_{\mu\nu\rho} = 0$. For the spin-4 field we use the same decomposition as in Eq.~\eqref{dec-spin4} with the identification $\varphi_{\mu\nu\rho\sigma} = \Phi_{\mu\nu\rho\sigma}$.

We now profit again from the fact that in two spatial dimensions any traceless symmetric tensor has only two independent components. Therefore, for every spatial traceless symmetric tensor of the form $Z_{a_1\cdots a_k}$ one can build a complex field $\mathcal Z_{[k]}$ out of its independent components as
\beq\label{complexfields}
Z_{a_1\cdots a_k}
\ \longrightarrow \
 \mathcal Z_{[k]}= Z_{\underset{k\,{\rm times}}{\,1\,1\,\cdots\,1}}
+ i\,
Z_{\underset{k-1\,{\rm times}}{\,1\,1\,\cdots\,1}\,2}.
\eeq
Using the prescription \eqref{complexfields}, we then define complex fields $\mS_{[k]}$ and $\mV_{[k-1]}$ associated to the fields $S_{a_1\cdots b_k}$ and $v_{a_1\cdots a_{k-1}}$ denoting the two spatial components of highest rank entering the decomposition of each field in the Lagrangian \eqref{Ls}. Similar complex fields can be introduced also for the components with less spatial indices (with the exception of the scalar ones). In analogy with what we discussed in the spin-2 case, the latter however always decouple in the limit $c \to \infty$.
As for spin two, before taking the non-relativistic limit we then introduce the field redefinition \eqref{V-U} for all fields $\mS_{[k]}$ and $\mV_{[k-1]}$. This, in turn, introduces new symmetric traceless tensors $Q_{a_1\cdots a_i}$ and $u_{a_1\cdots a_j}$, whose components are given by the real and imaginary parts of $\mQ_{[k]}$ and $\mU_{[k]}$ as indicated in \eqref{complexfields}.

After these redefinitions, the Lagrangian \eqref{Ls}, evaluated for $D=2+1$, becomes
\begin{widetext}
\beq\label{KandMintermsofNRfields}
\bal
\kappa^{-1} \mathcal L&= \sum_{k=1}^s 2^{k-2} \bigg[ im \dot \mQ_{[k]} \mQ_{[k]}^*
+ \left(\frac{m^2c^2}{2}(1-m_k)-mE_0\right)  |\mQ_{[k]}|^2
+ \frac{1}{2} \left(\frac{k}{2}-1\right) |\partial \mQ_{[k]}|^2 
\\
&
+
\frac{k}{2}\; m_k \,m^2|\mU_{[k-1]}|^2
+ \frac{im\,k}{2}
\mU_{[k-1]}\partial \mQ_{[k]}^*
+ \frac{c_kmc}{2}\bigg(im \,\mU_{[k-1]}^*  \mQ_{[k-1]}
-\mQ_{[k-1]} \partial \mQ_{[k]}^* 
\bigg) \bigg] +\rm{c.c.}
+ O(c^{-2}) 
\eal
\eeq
\end{widetext}
Notice that in order to eliminate the term diverging like $c^2$ in the contribution with $k = s$ in Eq.~\eqref{KandMintermsofNRfields}, the constant $m_s$ should take precisely the form \eqref{mk}. The contribution with $k = s$ contains another divergent term linear in $c$, which however cancels after integrating out $\mU_{[s-1]}$ from the action. 
Indeed, inspection of Eq.~\eqref{KandMintermsofNRfields} shows that the portion of the Lagrangian involving $\mU_{[s-1]}$ has the following structure: 
\beq\label{LVs}
\bal
\frac{2^{3-s}}{\kappa} \mathcal L_{\mU_{[s-1]}} & = i m\mU_{[s-1]}^* \bigg( c_s\,mc\, \mQ_{[s-1]}
-s \bar\partial \mQ_{[s]}
\bigg) \\
& + \frac{s\, m_s m^2}{2}|\mU_{[s-1]}|^2 + \rm{c.c.} + O(c^{-2}) 
\eal
\eeq
from which we find the following field equation for $\mU_{[s-1]}$ for large values of $c$:
\beq
\mU_{[s-1]}= -\frac{i}{ s\;m_s} \bigg( c_s\,c \mQ_{[s-1]}
-\frac{s}{m} \bar\partial \mQ_{[s]}
\bigg)
+ O(c^{-1}).
\eeq
Integrating out $\mU_{[s-1]}$ yields the effective Lagrangian
\beq\label{LeffVs}
 \mathcal L^{eff}_{\mU_{[s-1]}}=-
 \frac{2^{s-3} \kappa }{s\;m_s} 
 \bigg| c_s\,mc\, \mQ_{[s-1]}
-s \bar\partial \mQ_{[s]}
\bigg|^2 +\cdots,
\eeq
where $\cdots$ represents either terms that do not involve $\mQ_{[s]}$ or are subleading in $1/c$. Replacing \eqref{LVs} by \eqref{LeffVs} in the Lagrangian \eqref{KandMintermsofNRfields}, then cancels the term diverging linearly in $c$ coming from the contribution with $k = s$. A similar mechanism allows to cancel all divergences in $c$ by integrating out the fields $\mU_{[k]}$ recursively. Eventually, taking the limit $c\rightarrow \infty$ gives
\beq
\mathcal L= 2^{s-1}\kappa\;m\bigg[i \dot \mQ_{[s]} \mQ_{[s]}^*
-\frac1{2m}|\partial \mQ_{[s]}|^2
-E_0  |\mQ_{[s]}|^2
\bigg]+\dots,
\eeq
where $\mQ_{[s]}$ is decoupled from all the other fields in the Lagrangian. 

Even though we focused on the $s=4$ case, the previous derivation is written in a general form that allows one to generalize the result to arbitrary values of $s$. Moreover, as anticipated in the main text, using, e.g., that up to total derivatives $\epsilon_{mn}Q_{abcm}\dot Q_{abcn}=-8i\dot\mQ_{[4]} \mQ^*_{[4]}$, the Lagrangian for $\mQ_{[4]}$ takes the form in Eq.~\eqref{Landau3} with parameters $k'_1= -8\kappa m$, $k'_2= 4\kappa$ and $k'_3= 2\kappa m E_0$.

\bibliography{references}  

\end{document}